\documentclass[12pt]{article}
\usepackage{graphicx}
\usepackage{xcolor}
\usepackage{amsmath,slashed, amssymb}
\usepackage{fancyhdr}
\usepackage{hyperref}
\usepackage{mathtools}
\usepackage[titletoc]{appendix}
\numberwithin{equation}{section} 

\def\be{\begin{equation}}
\def\ee{\end{equation}}
\def\beq{\begin{equation}\begin{aligned}}
\def\eeq{\end{aligned}\end{equation}}

\newcommand{\pd}[2]{\frac{\partial #1}{\partial #2}}

\newcommand{\dd}{\mathrm{d}}

\newcommand{\GeV}{\text{GeV}}

\newcommand{\gsim}{\lower.7ex\hbox{$\;\stackrel{\textstyle>}{\sim}\;$}}
\newcommand{\lsim}{\lower.7ex\hbox{$\;\stackrel{\textstyle<}{\sim}\;$}}

\def\centeron#1#2{{\setbox0=\hbox{#1}\setbox1=\hbox{#2}\ifdim
\wd1\rangle\wd0\kern.5\wd1\kern-.5\wd0\fi
\copy0\kern-.5\wd0\kern-.5\wd1\copy1\ifdim\wd0\rangle\wd1
\kern.5\wd0\kern-.5\wd1\fi}}
\def\ltap{\;\centeron{\raise.35ex\hbox{$\langle$}}{\lower.65ex\hbox{$\sim$}}\;}
\def\gtap{\;\centeron{\raise.35ex\hbox{$\rangle$}}{\lower.65ex\hbox{$\sim$}}\;}
\def\gsim{\mathrel{\gtap}}
\def\lsim{\mathrel{\ltap}}

\newcommand{\newc}{\newcommand}
\newc{\qbar}{{\overline q}}
\newc{\Kahler}{Kahler }
\newc{\deltaGS}{\delta_{\rm GS}}

\newcommand{\AG}[1]{\textbf{\textcolor{red}{[AG: #1]}}} 
\newcommand{\MD}[1]{\textbf{\textcolor{blue}{[MD: #1]}}}

\begin{document}
\begin{titlepage}
\begin{flushright}
{\large SCIPP 20/01\\}
\end{flushright}

\vskip 1.2cm

\begin{center}

{\LARGE\bf Low Energy Effective Theory for Axion Strings}

\vskip 1.4cm

{\large Michael Dine$^{(a)}$}
\\
\vskip 0.4cm
{\it $^{(a)}$Santa Cruz Institute for Particle Physics and
\\ Department of Physics, University of California at Santa Cruz \\
     Santa Cruz, CA 95064.} \\
\vspace{0.3cm}

\end{center}
\vskip 4pt


\begin{abstract}
We consider aspects of the low energy (classical) effective field theory description of global cosmic strings.  In the non-relativistic limit, we study the extent to which these are described by the combination of the Nambu-Gotto and
Kalb-Ramond actions.  While in a formal sense this is the case for infinitely long strings, for more realistic situations, the collective
coordinates of the system do not provide a complete description for more than brief intervals; 
as time evolves it becomes necessary to include low momentum goldstone excitations as well. 

\end{abstract}

\end{titlepage}

\section{Introduction}

There has been ongoing discussion for almost four decades of the possible role of axion strings in axion dark matter production (see, e.g., \cite{Davis:1986xc,Davis:1989nj,Battye:1993jv,Battye:1994au, dabholkar,vilenkin}.  There are a range of analyses, some analytical, many numerical
simulations, with the latter quite large and sophisticated in recent years\cite{villadoro,guymoore,Fleury:2015aca,sikivie1,sikivie2,sikivie3,safdi}.  In some cases,  these simulations appear to point to a parameterically
enhanced production of axion dark matter due to cosmic strings;
in other cases, not so much so.  There is a large parameter in the problem,  the {\it infrared sensitive} value of the string tension divided by $f_a^2$. Assuming that the
Hubble constant provides the infrared cutoff, this parameter is typically taken to be:
\beq
\xi \equiv \log(f_a/H_{qcd}) \sim 70.\eeq
  If there is of order one string
per horizon, and if most of the energy in the axion field surrounding axion strings
were converted into very low momentum axions, there would be an enhancement  of the dark matter axion density by such a factor over naive, textbook\cite{kalbramond}
considerations.  This could have significant consequences for cosmology and for the ability of experiments such as ADMX to detect the dark
matter.  In \cite{dfgp}, qualitative arguments were put forth that there is no such enhancement.  This work {\it did} enumerate a number of challenges for analytic dark
matter estimates at the order one level, which
simulations have the potential to resolve.

In this paper, we will look more closely at the effective theory of axion strings, asking whether one might expect enhancements of axion emission at low momentum by powers of $\xi$.
In effect, we are revisiting the older literature on these objects.  Crucial will be the role of the infrared in axion production.  Simulations, when sufficiently accurate, generally consider
all degrees of freedom simultaneously.  They do not exploit low energy actions  for collective coordinates and low momentum axions of the type we will consider here.    It would be interesting
to perform simulations in the low energy effective theory, and/or to match the results of simulations to these theories.

As opposed to Nielsen-Olesen, gauge strings\cite{nielsenolesen}, strings associated with the breaking of a global symmetry have some problematic features.  In particular, if one tries to define a tension, one finds that it is infrared
divergent.  As with all infrared divergences, this is a signal, as stressed in \cite{dfgp}, that one must include an enlarged set of degrees of freedom in describing the system\footnote{It is perhaps worth
recalling how this is realized in QED.  There, the three point function is infrared divergent.  It cannot be used in an effective field theory by itself to compute scattering amplitudes; it is necessary to 
consider emission of extremely low energy photons as well.}.   In the case
of global cosmic strings, this means that one can't consider the system of string collective coordinates by themselves.  Instead, it is necessary to consider an effective action involving these coordinates and axions with momentum below some cutoff scale.  For an infinitely long string, the cutoff should be chosen so as to avoid problems
of causality.  Implementing such a cutoff might be complicated.  For a closed string of circumference of order $R$, $R$ itself cuts off infrared divergences, and ideally one should choose the (momentum space) cutoff large compared to $R^{-1}$.  For long, parallel string anti-string segments separated by a distance $b$, $b$ acts as a cutoff.  Such configurations are likely more representative than the infinite string of the actual cosmological situation.  In this note, we will look at the effective action for this
system, particularly
in the non-relativistic limit in which the description of solitons becomes simple.  We will see that for a long straight string, with little excitation of non-zero modes, the Nambu-Goto action, supplemented
with the Kalb-Ramond coupling\cite{kalbramond} of the axion to the string (NGKR), in a certain formal sense, captures the dynamics.  But
for closed strings or string-antistring segments the NGKR description, by itself, fails to correctly capture the dynamics of the system. The problem is that while, at one instant,
the system might be well described by its collective coordinates without axion excitations, this will typically not be true a short time later.  This is in sharp contrast to infrared
{\it insensitive} systems like magnetic monopoles and gauge strings in four dimensions.  Said another way, for most of its history, the string, as an entity, is not the simple object considered in many analyses. Note, here, we are speaking of the classical (as opposed to quantum) theory of these strings.  This
will be our focus throughout this paper.

The rest of this paper is organized as follows.  In the next section, we briefly review well known aspects of the collective coordinate method for solitons, both particles
and strings.  For systems such as monopoles and Nielsen-Olsen strings, we explain the
sense in which the collective coordinates provide a complete description of the system, in uniform motion, for all times.  We consider the collective coordinate analysis for an infinitely long, straight global string in section
\ref{fieldtheory}.  We derive, formally, the non-relativistic Nambu-Gotto action for the collective coordinates, with its infrared divergent tension, and in section \ref{kalbramondaction} we will see that the Kalb-Ramond coupling does formally describe the axion couplings to the collective modes. We then consider, in section \ref{closedstrings}, the description of closed string systems.   We note that, related to the infrared divergence of the infinite string, there is not a unique definition of collective coordinates.  It is then not surprising, as we elucidate, that the NGKR description
of such systems is inadequate.  In section \ref{conclusions}, we summarize and describe our expectations for the spectrum of axion radiation emitted
by the sorts of closed string systems one might expect to encounter in early universe cosmology and in simulations of such cosmologies.

\section{Collective Coordinate Review}
\label{collectivecoordinates}

The utility of the collective coordinate method follows from elementary considerations.  Consider, for example, a kink in a translationally invariant
theory in two dimensions, described by a classical solution, $\Phi_{cl}(x)$
\beq
\Phi(x,t) = \Phi_{cl}(x-X(t)) + \delta \Phi.
\label{phicollectivea}
\eeq
$\delta \Phi$ describes the fluctuations about the classical solution.   We can obtain an action for the degrees of freedom $X$ and $\delta \Phi$
by substituting \ref{phicollectivea} into the field theory action.
There is a kinetic term for $X$ of the form
\beq
{\cal L}(X,\delta \Phi) = {1 \over 2} M \dot X^2 + \dots
\eeq
and higher order terms in $\dot X$.  Because of translational invariance, there are no terms involving $X$ without derivatives.
In addition, precisely because $\Phi_{cl}$ {\it is} a solution of the classical equations, when we work out the action for $X$, there are no terms in the action linear in $\delta \Phi$ for a kink
in uniform motion.
So if one has a soliton in uniform motion at time $t=t_0$, say, with $\Delta \Phi =0$, $\delta \Phi$ remains zero for all time.

The same is true for other collective coordinates associated with symmetries of solutions, such as translational and dyonic excitations of monopoles.
Infinite gauge strings in four dimensions have two collective coordinates, the transverse coordinates of the string solution, $\vec X_\bot$.  Calling
$z$ the coordinate along the string, and writing the field in the presence of the string as
\beq
\Phi(\vec x_\perp,z,t) = \Phi(\vec x_\perp - \vec X(z,t)) + \delta \Phi,
\eeq
the action for these collective coordinates has the form:
\beq
\int d^2 \sigma {\cal L}(X,\delta \Phi) =\int d^2 \sigma {1 \over 2} T \left ({\partial \vec X \over \partial \sigma_\alpha} \right)^2 + \dots
\eeq 
Here $\sigma =(z,t)$; this is the non-relativistic limit of the usual Nambu-Gotto action.  
Again, there is no term linear in $\delta \Phi$.  So, for a string executing uniform motion, starting with $\delta \Phi=0$ at some time, $\delta \Phi$ remains zero.  All of this follows
from the translational invariance of the theory. The Nambu-Gotto description remains good if the string is curved, with curvature small compared to the characteristic
scales of the system.

This simple observation is the origin of the power of the collective coordinate method for such systems.  It will not be the case for strings arising from the breaking of global symmetries in four dimensions (or vortices in $2+1$ dimensions).  This is precisely because of the infrared divergences.  The effects which cut off the divergence,
such as finite circumference, will spoil the collective coordinate story, in the sense that the system as described by the collective coordinates alone
does not obey the equations of motion.  This is closely tied to the need to include other degrees of freedom
in the effective action.  The infrared divergence yields a dependence on the breaking scale of the symmetry.
There will be a potential for the collective coordinates
and coupling of the collective coordinates to $\delta \Phi$.

\section{The Effective Action For the Collective Coordinates for a Long Global String}
\label{fieldtheory}

Consider a theory of a complex scalar, $\Phi$, with action symmetric under $\Phi \rightarrow e^{i \alpha} \Phi$:
\beq
S= \int d^4 x \left ( \vert \partial_\mu \Phi \vert^2 + m^2 \vert \Phi\vert^2 - {\lambda \over 2} \vert \Phi \vert^4 \right ).
\eeq 
For this action, 
\beq
\langle \vert \Phi \vert\rangle \equiv f_a \equiv v = {m^2 \over \lambda}.
\eeq

This action admits static string solutions.  Taking the string to lie along the $z$ axis, with coordinates $(x,y,z) \equiv (\vec x_\bot, z) = (\rho,\theta,z)$,
\beq
\Phi(t,z,\vec x_\bot) = \Phi_{cl}(\vec x_\bot) = f(\rho) e^{i\theta}.
\eeq
The function $f(\rho)$ has the properties:
\beq
f(\rho) \rightarrow 0~{\rm as}~\rho \rightarrow 0;~~~f(\rho) \rightarrow f_a (1 - {1 \over f_a \rho^2}) ~{\rm as} ~\rho \rightarrow \infty.
\eeq

We first obtain the effective action for the collective coordinates for the string.
Note, first that if we calculate the tension, thought of as the energy per unit length of the string, the result is infrared divergent:
\beq
T = \int d^2 x_\bot \left ( \vert \vec \nabla \Phi\vert^2 + V(\Phi) - V_0 \right ) \approx 2 \pi f_a^2\int d\rho {1 \over \rho^2}  = 2 \pi \log(\Lambda/\mu).
\eeq
Here $\Lambda$ is an ultraviolet cutoff, of order the core size; we'll take $\Lambda = f_a$.  $\mu$ is an infrared cutoff.  If we have a closed string
with radius $R$, $\mu = R^{-1}$; if we have a parallel string and antistring separated by a distance $b$, $\mu = b^{-1}$.

We can see more directly that $T$ is the tension of the string by
allowing slow variation in the transverse directions, introducing collective coordinates $X^i(z,t),~i=1,2$.  We also allow variation of the phase of $\Phi$ from
$i \theta$, i.e. a spatially and time-dependent axion field:
\beq
\Phi(t,z,\vec x_\bot) = \Phi_{cl}( \vec X(t,z) -\vec x_\bot) e^{i \delta a(t,z,\vec x_\bot)}.
\label{collectivecoordinates}
\eeq
Now plug this expression into the terms in the action:
\beq
S= \int dt dz d^2 x_\bot \left (\vert \partial_t \Phi \vert^2 - \vert \partial_z \Phi \vert^2 \right)
\eeq
$$~~~\approx \int dt dz \left ((\partial_t X^i)^2 - (\partial_z X^i)^2 \right ) \int d^2 x_\bot \vert \vec \nabla \Phi_{cl}(x_\bot) \vert^2$$
$$~~~~~+ \int d^4x f_a^2 \left ( \partial_t X^i \partial_i ~\theta \partial_t \delta a- \partial_z X^i \partial_i ~\theta \partial_z \delta a \right ).$$
$$~~~~\equiv I_{transverse} + I_{\delta a}$$
This is valid for distances from the string small compared to the wavelength of the disturbance.  In this regime, the variation of the coordinate is roughly constant, and the light transit time
can be neglected.  The first term can be rewritten as
\beq
I_{transverse} = T \int {dz dt} \left (\left ({\partial \vec X_{\bot}(t,z) \over \partial t}\right )^2 -\left ({\partial \vec X_{\bot}(t,z) \over \partial z}\right)^2 \right )
\label{nambugototerm}
\eeq
with $T$ given by our earlier expression.  The infrared divergence appears again.  
In the second term,$\partial_i \theta = {\epsilon_{ij} x^j \over \vert \vec x_\bot \vert^2}.$ 
so
\beq
I_{\delta a} = \int d^4 x \epsilon_{ij} {x^j \over \vert \vec x_\bot \vert^2}  f_a^2 \left ( \partial_t X^i  \partial_t \delta a- \partial_z X^i  ~\theta \partial_z \delta a \right ).
\label{krterm}
\eeq

Equation \ref{nambugototerm} is, again, the non-relativistic limit of the Nambu-Goto action, but with an infrared divergent tension.  Before considering how this divergence may be tamed,
we can ask about the coupling to $\delta a$.  As we now demonstrate (and was remarked in \cite{dfgp}), equation \ref{krterm} is of the form of the Kalb-Ramond action\cite{kalbramond}.

\section{Appearance of the KR term for a Long, Slowly Moving String}
\label{kalbramondaction}

In this section, we observe formally that for an infinitely long, straight string, the KR action {\it does} describe the coupling of the string
collective coordinates to axion radiation for a slowly moving string.
The $a$ configuration defining the string changes as $X^i$, the string collective coordinates, change.  If the collective coordinates change slowly enough, we might expect $a$ to be, at each instant, nearly in its lowest energy (the $\nabla^2 a =0$) configuration.  The slight differences will correspond to radiation.   This is encoded in our collective
coordinate analysis, and we will see it is reproduced by the KR action.

\subsection{The Circulating Axion Field}

We first express the axion field surrounding a static string in terms of the antisymmetric tensor field.
Consider an infinite static string along the $z$ direction, i.e. $X^0 =t$, $X^z = z$.  The KR action leads to the equation:
\beq
\Box B_{0z} = f_{KR}.
\eeq
(We will fix the constant, $f_{KR}$ in a moment.)   So, taking $B_{0z}$ to be $z$ and $t$ independent,
\beq
B_{0z} = G(\vec x_\bot) f_{KR}
\eeq  
where $G(\vec x_\bot)$ is the two dimensional Green's function,
\beq
G(\vec x_\bot) = {1 \over 2 \pi} \log(\vert \vec x_\bot \vert).
\eeq
From this, we can construct
\beq
H_{0zi} = \partial_i B_{0z} = \epsilon_{0zij} \partial_j a
\eeq
so
\beq
\partial_j a = \epsilon_{ij} \partial_i f_{KR} G = \epsilon_{ij} \partial_j f_{KR} {x^i \over \vert \vec x_\bot \vert^2}
\eeq
which is $f_{KR} \partial_j \theta$, the polar angle in cylindrical coordinates.  So the static configuration is as expected.  Note that this is the minimal
energy configuration for fixed $B$ ($a$), with the boundary condition of unit winding. 

\subsection{Axion Emission From the String}

Now consider axion emission from a moving string.  Consider the term in the action
\beq
\int d^4 x{\cal L} \sim \int d^4 x \left ( H_{zij}^2 -H_{0ij}^2 \right ).
\eeq
Noting, for example, that
\beq
H_{zij}(t,z,\vec x_\bot) = \partial_0 a + \epsilon_{ij} \partial_i G^{(2)}(\vec x_\bot) {\partial X^j \over \partial t}  
\eeq
yields for the action a sum of terms, ${\cal L} ={\cal L}_{X^i}+ {\cal L}_{a-X^i} $, where:
\beq
{\cal L}_{X^i}=\int d^4 x{\cal L} = \int dz~dt x \left (({dX^i \over dt})^2 -({dX^i \over dz})^2 \right ) \int d^2 x_\bot  \left ( {1 \over x_\bot^2} \right )  
\eeq
and
\beq
{\cal L}_{a-X^i} = \int dz dt ~d^2 x_\bot \left (\partial_0 a {d X^i \over dt} - \partial_z a {dX^i \over dz} \right ) \epsilon_{ij} {x^j \over x_\bot^2} .
\eeq
The first expression is identical to the action for the transverse fluctuations of the string found in the field theory; the second is the
axion-string collective coordinate coupling found there.

So formally, the collective coordinates and the axion perturbation of the solitonic string in field theory are described by the Nambu-Gotto-Kalb-Ramond (NGKR)
action.  We say formally both because of the infrared divergent tension and because of issues of causality:  the axion field changes everywhere in response to
transverse string displacements.  These issues should be addressed by systems of closed strings, such as long, parallel strings and antistrings or circular strings, where string separations and/or radii will act as an
infrared cutoff.  We consider these issues in the next section.  As we anticipated earlier, the fact that these configurations are not static solutions of the
equations of motion limits the applicability of the collective coordinate description.

\section{Systems of Closed Strings}
\label{closedstrings}

In \cite{dfgp}, two limits of string motion were considered, referred to as ``adiabatic" and ``sudden".  These were considered in a rather heuristic way.  The adiabatic limit corresponds to slow motion of the string, and, in this limit, as the strings moves and the axion field surrounding them changes, the energy change in the axion field 
is compensated by the change in the kinetic energy of the collective coordinate.  The sudden limit corresponds to fast motion of the string.  In this case, most of the energy
stored in the axion field is converted into freely moving axions.  If this picture is correct, than in the case of adiabatic motion, the system should be well described by
its collective coordinates; this is not true of the sudden case.  In this section we will understand this picture in more detail in terms of the equations of motion of the system.

Because of the infrared issues associated with the axion field, equation \ref{collectivecoordinates} is not really sensible for an infinite string; it also implies an instantaneous
change in the axion field throughout space.  For a closed system, e.g. a closed, roughly circular string, or alternatively for a long, parallel string and antistring,
this potentially makes more sense, as the infrared divergences cancel.   In this case, however,
we have to consider what we mean by $\Phi_{cl}$, since there are not static solutions of the equations of motion of this type.  We might take this to be, for example, the minimal energy static (instantaneously) configuration with
specified location of the string core.  Then we can make the substitution of equation \ref{collectivecoordinates} and obtain an action for these coordinates.  
{\it  But the question of what configuration our collective coordinates describe does not have a unique answer.}  We have no reason to expect that as the system
evolves, even if it starts in one such configuration, say, again, the axion configuration of lowest energy, it will remain in the corresponding lowest energy configuration.

For a long string-antistring pair, separated by a distance $b$ and moving slowly, the field configuration of lowest energy, far from the string core, is just the
superposition of two static solutions.  The corresponding field
configuration is, far from the string core:
\beq
\Phi = f_a e^{i{a(x,t) + \delta a \over f_a}} 
\label{phicollective}
\eeq
where
\beq
a(x,t) =  a_0(x_\bot- {b(t) \hat y \over 2})-a_0(x_\bot+ {b(t) \hat y \over 2}) .
\eeq
Here $a_0(\vec x_0)$ is the axion configuration around a single string.  We now show that with initial conditions
\beq
b(0) = b_0;~\dot b(0) = 0
\label{initialconditions}
\eeq
this is {\it almost} a solution of the equations of motion, for short times.

Substituting $\Phi$ from equation \ref{phicollective} into the field theory action,
yields an action for $b$:
\beq
L\approx f_a^2 \ell (\dot b^2 \log(f_a b) + \log(f_a b)).
\eeq
as well as a coupling of the collective mode, $b$, to the axion, $\delta a$:
\beq
\int d^3 x~ dt~ \left(\dot b {\partial a\over \partial b} \partial_t \delta a- {\partial a \over \partial x^i} {\partial \delta a \over \partial x^i}\right )
\eeq

 We can integrate by parts in the first term with respect to time.  Provided that $\dot b$ is small, the source for $\delta a$ is then proportional to $\ddot b ={1 \over b \log(f_a b)}$.  The {\it energy} stored in $\delta a$ is suppressed, at early times, by $1/\over \log(f_a b)$ relative to the kinetic energy in $b$.  But we can make a stronger statement.   The system can be described
as approximately adiabatic\cite{dfgp}.  Note that the source for the
mode
\beq
\delta a = C(t) {\partial a \over \partial b}
\eeq
vanishes at early times as a consequence of the equation of motion for $b$.
To see this, note that, the kinetic term,  after integration by parts, neglecting $\dot b$, yields:
\beq
\ddot b C \int d^2 x_\bot ({\partial a \over \partial b})^2 
\eeq
which is just $C$ times the second derivative term in the equation of motion for $b$,
while from the integration over the $\vec \nabla_\bot^2$ term one has:
\beq
C {\partial \over \partial b} \int d^2 x_\bot (\vec \nabla a)^2 =C {\partial V(b) \over \partial b}.
\eeq

So, provided $\dot b\approx 0$,  the field of equation \ref{phicollective} is {\it almost} a solution of the equations of motion.  But as $\dot b$ grows, $C$ is significantly sourced.  In particular, the $\dot b^2$ terms dominate once $\dot b > \log (f_a b)^{-1/2}.$  At this point the
adiabatic approximation is breaking down.

This is quite general.  For circular motion, one instead studies a classical solution with a circular core of radius $R$.  One can again start with a static configuration, minimizing
the axion field energy as a function of $R$.  Again the mode:
\beq
\delta a = C(t) {\partial a \over \partial R}
\eeq 
is not sourced initially as a consequence of the equations of motion for $R$.  Now, however, the system quickly becomes relativistic, and $C$ is driven away from zero.
Correspondingly, much of the field energy is converted into radiation, rather than kinetic energy of $R$.

Loosely speaking, treatments of radiation utilizing the Kalb-Ramond action require a careful treatment of ``back reaction". Once one enters the non-adiabatic regime, most of the energy of the field is converted to radiation rather than kinetic energy of $b$.  To fully account for this requires solving the full set of equations.  It is reassuring, at least, that both regimes can be analyzed, for the string-antistring case, in a non-relativistic limit.  Circular strings become relativistic more quickly, and these back-reaction effects will be pronounced early on.

We conclude from this that initially for a long parallel string and antistring, if the system starts at rest, the bulk of the energy in the axion field is converted into kinetic energy of $b$.  Little radiation is produced.  As $b$ decreases and the velocity becomes relativistic, production of axions increases.  Note, in particular, that one produces the mode $C$.  This corresponds to the fact that, as the velocity increases,
this motion is less and less an approximate solution of the equations of motion.  All of this is consistent with the expectation that the energy in the axion field at a length scale $b(t)$ is converted into radiation of wavelength $b(t)$ except, possibly, at early stages of the motion, where long wavelength axions are suppressed if the velocity, $\dot b$, is
small.

\section{Conclusions}
\label{conclusions}

Global strings are peculiar objects.  In isolation, their tensions are infrared divergent.  This indicates that they are not really objects at all, in that they cannot, except possibly for brief periods, be described
simply in terms of their collective coordinates.  We have seen this explicitly for long, parallel strings and antistrings, and for closed strings.  For these systems,
the infrared divergences cancel, but static configurations are not solutions of the equations of motion.  We have given a plausible definition of
the collective coordinates for such systems.  But as these systems move, even if initially the collective coordinates provide a good description, the description
quickly becomes incomplete, with a substantial (typically order one or larger) fraction of the energy in excitations.  In this sense, the Nambu-Gato-Kalb-Ramond
description is inadequate.  As signaled by the infrared divergence, it is necessary to consider degrees of freedom beyond the string collective coordinates.

As we have recalled, in \cite{dfgp}, at a heuristic level two limiting cases of axion radiation by strings were considered for a long, parallel string and antistring.  One, referred to as adiabatic, involved slow motion of strings, with the axion field adjusting at each instant to the configuration of the string core.  In this limit, it was argued, the energy of the axion field would largely be converted to kinetic energy
of the collective coordinate $b$.  In this paper, we have seen how this is realized explicitly at the level of the equations of the effective field theory, in the limit that $\dot b$ is small.
A second limit, referred to as sudden, involved rapid motion of the strings, with most of the energy in the axion field of the strings converted to axions.  Again, we have seen
this at the level of the field theory equations when $\dot b$ is not small.  In the small $\dot b$, adiabatic case, few axions are produced,  In the large $\dot b$ limit, most of the energy in the axion field surrounding the strings is converted to axions, but these have a spread in wavelength, with typical wavelengths at time $t$ of order
the string separation at time $t$.  From this, we see that in the sudden limit, the system is not well described by the collective coordinates alone.
We also see that there is no reason to expect a lowlogarithmic enhancement of the low
energy axion density; rather the energy should be distributed uniformly (on a log scale) over wave numbers between the Hubble scale and the scale $f_a$.  These observations support the picture, put forward in \cite{dfgp}, that cosmic strings are unlikely to yield a parameterically enhanced contribution to
the population of low momentum axions, and thus to an enhanced axion dark matter density.  Recent simulations with substantial reach in $\xi$\cite{safdi} appear consistent with these expectations.

\vskip 1cm
\noindent
\noindent
{\bf Acknowledgements:}  I am grateful for collaboration and conversations with Nicolas Fernandez, Akshay Ghalsasiand Hiren Patel.
This work was supported in part by the U.S. Department of Energy grant number DE-FG02-04ER41286.

\bibliography{axion_cosmic_strings}{}
\bibliographystyle{utphys}

\end{document}